\documentclass[prd,showpacs,preprintnumbers,floatfix,twocolumn]{revtex4}
\usepackage{amsmath}
\usepackage{graphicx}
\begin{document}
\title{Quantum Entanglement of Electromagnetic Field in Non-inertial Reference Frames}
\date{\today }
\author{Yi Ling}
\email{yling@ncu.edu.cn} \affiliation{Center for Gravity and
Relativistic Astrophysics, Department of Physics, Nanchang
University, Nanchang, 330047, China\\
CCAST (World Laboratory), P.O. Box 8730, Beijing,
   100080, China}
\author{Song He}
 \affiliation{Institute of Theoretical Physics, School of Physics, Peking
University, Beijing, 100871, China}
\author{Weigang Qiu}
\affiliation{Department of Physics, Huzhou Teachers College, Huzhou,
313000, China}
\author{Hongbao Zhang}
    \affiliation{Department of Astronomy, Beijing Normal University, Beijing, 100875,
    China\\
    Department of Physics,
Beijing Normal University, Beijing, 100875, China\\
 CCAST (World
Laboratory), P.O. Box 8730, Beijing,
   100080, China}
\begin{abstract}
Recently relativistic quantum information has received considerable
attention due to its theoretical importance and practical
application. Especially, quantum entanglement in non-inertial
reference frames has been studied for scalar and Dirac fields. As a
further step along this line, we here shall investigate quantum
entanglement of electromagnetic field in non-inertial reference
frames. In particular, the entanglement of photon helicity entangled
state is extensively analyzed. Interestingly, the resultant
logarithmic negativity and mutual information remain the same as
those for inertial reference frames, which is completely different
from that previously obtained for the particle number entangled
state.
\end{abstract}

 \maketitle

\section{Introduction}
Quantum entanglement is both the central concept and the major
resource in quantum information science such as quantum
teleportation and quantum computation\cite{BEZ}. In recent years,
tremendous progress has been made in the research on quantum
entanglement: not only have remarkable results been obtained in this
field, but also important techniques been applied to various
circumstances\cite{PV}.

Especially, considerable effort has been expended on the
investigation of quantum entanglement in the relativistic framework
recently\cite{AET,PT,Shi}. A key issue in this intriguing and active
research direction is whether quantum entanglement is
observer-dependent. It has been shown that quantum entanglement
remains invariant between inertial observers with relative motion in
flat spacetime although the entanglement between some degrees of
freedom can be transferred to others\cite{Peres1,AM1,GA,He}.
However, for scalar and Dirac fields, the degradation of
entanglement will occur from the perspective of a uniformly
accelerated observer, which essentially originates from the fact
that the event horizon appears and Unruh effect results in a loss of
information for the non-inertial observer\cite{AM2,AM3,FM,AFMT}.

As a further step along this line, this paper will provide an
analysis of quantum entanglement of electromagnetic field in
non-inertial reference frames. In particular, we here choose the
photon helicity entangled state
$\frac{1}{\sqrt{2}}(|\uparrow\rangle_A|\downarrow\rangle_B+|\downarrow\rangle_A|\uparrow\rangle_B)$
rather than the particle number entangled state
$\frac{1}{\sqrt{2}}(|0\rangle_A|0\rangle_B+|1\rangle_A|1\rangle_B)$
in an inertial reference frame as our main point for investigation
of quantum entanglement in non-inertial reference frames, where A
and B represent an inertial observer Alice, and a uniformly
accelerated observer Bob respectively,  as is illustrated in
FIG.\ref{AliceBob}. It thus makes the present work acquire much
interest and significance: the former entangled state seems to be
more popular in quantum information science, but previous work only
restricts within the latter setting\cite{AM2,AM3,FM,AFMT}. In
addition, the result obtained here shows that although Bob is forced
to trace over a causally disconnected region of spacetime that he
can not access due to his acceleration, which also leads his
description of the helicity entangled state to take the form of a
mixed state; the corresponding logarithmic negativity and mutual
information both remain invariant against the acceleration of Bob.
Therefore our result is of remarkable novelty: it is completely
different from those obtained for the case of the particle number
entangled state, where the degradation of entanglement is dependent
on the acceleration of observer, namely, the larger the
acceleration, the larger the degradation\cite{AM2,AM3,FM,AFMT}.

The paper is organized as follows. In the next section, we shall
briefly review the four disconnected sectors in Minkowski spacetime
and the accelerated observers in Rindler spacetime. In the
subsequent section, introducing the two sets of expansion bases for
quantizing the electromagnetic field in Minkowski spacetime, we have
developed the relationship between the corresponding annihilation
and creation operators in Minkowski spacetime. In Section \ref{QE},
we shall analyze quantum entanglement of electromagnetic field in
non-inertial reference frames, especially for the photon helicity
entangled state . Conclusions and discussions are presented in the
last section.

System of natural units are adopted: $\hbar=c=1$. In addition, the
metric signature takes $(+,-,-,-)$, and the Lorentz gauge condition
$\nabla_aA^a=0$ is imposed onto the electromagnetic potential in
flat spacetime, where Maxwell equation reads
\begin{equation}
\nabla_a\nabla^aA_b=0.\label{Maxwell}
\end{equation}
Moreover, the well known inner product is reduced to
\begin{equation}
(A,A')=i\int_\Sigma[\nabla^a\bar{A}^b)A'_b-\bar{A}_b\nabla^aA'^b]\epsilon_{acde},\label{inner}
\end{equation}
which is gauge invariant and independent of the choice of Cauchy
surface $\Sigma$\cite{Moretti,HQZ}.
\section{Accelerated Observers in Minkowski Spacetime}
\begin{figure}
  \includegraphics[width=2.5inch]{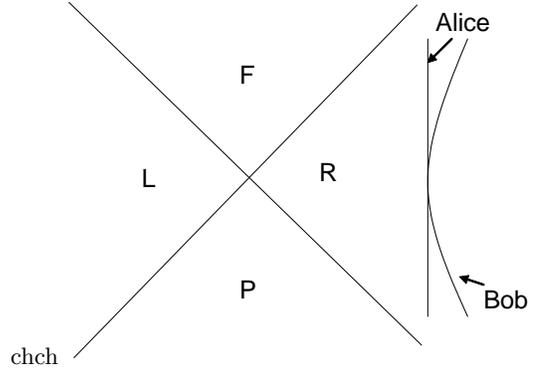}\\
  \caption{The four disconnected patches in Minkowski spacetime with an inertial observer Alice and a uniformly accelerated observer Bob constrained in $R$ sector. }\label{AliceBob}
\end{figure}

Start from Minkowski spacetime
\begin{equation}
ds^2=dt^2-dx^2-dy^2-dz^2.
\end{equation}
As is shown in FIG. \ref{AliceBob}, we perform the coordinate
transformations for the four disconnected sectors in Minkowski
spacetime, respectively, i.e.,
\begin{eqnarray}
&R&\nonumber\\
t=\rho\sinh\tau,&&x=\rho\cosh\tau,\nonumber\\
\rho=\sqrt{x^2-t^2},&&\tau=\tanh^{-1}(\frac{t}{x}),
\end{eqnarray}
\begin{eqnarray}
&L&\nonumber\\
t=\rho\sinh\tau,&&x=\rho\cosh\tau,\nonumber\\
\rho=-\sqrt{x^2-t^2},&&\tau=\tanh^{-1}(\frac{t}{x}),
\end{eqnarray}
\begin{eqnarray}
&F&\nonumber\\
t=\rho\cosh\tau,&&x=\rho\sinh\tau,\nonumber\\
\rho=\sqrt{t^2-x^2},&&\tau=\tanh^{-1}(\frac{x}{t}),
\end{eqnarray}
\begin{eqnarray}
&P&\nonumber\\
t=\rho\cosh\tau,&&x=\rho\sinh\tau\nonumber\\
\rho=-\sqrt{t^2-x^2},&&\tau=\tanh^{-1}(\frac{x}{t}).
\end{eqnarray}
In particular, the $R$($L$) sector, viewed as a spacetime in its own
right, is also called $R$($L$) Rindler spacetime, where the metric
reads
\begin{equation}
ds^2=\rho^2d\tau^2-d\rho^2-dy^2-dz^2,
\end{equation}
and the integral curves of boost Killing field
$(\frac{\partial}{\partial\tau})^a$ correspond to the worldlines of
accelerated observers with proper time $\rho\tau$ and acceleration
$\frac{1}{\rho}$.
\section{Quantum Electromagnetic Field in Minkowski Spacetime}
As is well known, the quantum fields can be expanded in terms of
various bases, but the corresponding vacua may be completely
different. For the quantum electromagnetic field in Minkowski
spacetime, we firstly choose the expansion basis as
\begin{eqnarray}
&&A_\mu(\omega\in R,p_y\in R,p_z\in R,
s=\pm 1)=\nonumber\\
&&\frac{1}{8\pi^2p_\bot}[(0,0,p_z\phi,-p_y\phi)+s(\partial_x\phi,\partial_t\phi,0,0)],
\end{eqnarray}
where $p_\bot=\sqrt{p_y^2+p_z^2}$, and
\begin{equation}
\phi=\int_{-\infty}^\infty d\lambda
e^{(-i\omega\lambda-ip_\bot\cosh\lambda t+ip_\bot\sinh\lambda
x+ip_yy+ip_zz)}
\end{equation}
satisfies Klein-Gordon equation in Minkowski spacetime, with
$\omega$ a dimensionless parameter\cite{Moretti,Colosi}.

It is easy to check that $A_\mu(\omega,p_y,p_z,s)$ is the
simultaneous eigensolution of boost, transverse momentum, and
helicity operators with the corresponding eigenvalues $\{\omega,p_y,
p_z, s\}$ in Minkowski spacetime\cite{HQZ,Ashtekar}. Furthermore, it
is orthonormal with respect to the inner product (\ref{inner}),
i.e.,
\begin{eqnarray}
&&(A(\omega,p_y,p_z,s),A(\omega',p'_y,p'_z,s'))=\nonumber\\
&&\delta(\omega-\omega')\delta(p_y-p'_y)\delta(p_z-p'_z)\delta_{ss'}.
\end{eqnarray}
Thus in terms of this basis, the quantum electromagnetic field can
be expanded as
\begin{eqnarray}
\hat{A}_\mu=&&\int_{-\infty}^\infty d\omega\int_{-\infty}^\infty
dp_y\int_{-\infty}^\infty dp_z\sum_{s=\pm1}\nonumber\\
&&[c(\omega,p_y,p_z,s)A_\mu(\omega,p_y,p_z,s)\nonumber\\
&&+c^\dagger(\omega,p_y,p_z,s)\bar{A}_\mu(\omega,p_y,p_z,s)],
\end{eqnarray}
where $c$ and $c^\dagger$ are the corresponding annihilation and
creation operators, respectively, adjoint to each other, and
satisfying the following commutation relations
\begin{equation}
[c(\omega,p_y,p_z,s),c(\omega',p'_y,p'_z,s')]=0,
\end{equation}
\begin{equation}
[c^\dagger(\omega,p_y,p_z,s),c^\dagger(\omega',p'_y,p'_z,s')]=0,
\end{equation}
\begin{eqnarray}
&&[c(\omega,p_y,p_z,s),c^\dagger(\omega',p'_y,p'_z,s')]=\nonumber\\
&&\delta(\omega-\omega')\delta(p_y-p'_y)\delta(p_z-p'_z)\delta_{ss'}.
\end{eqnarray}

Next we can also employ Unruh expansion basis for the quantum
electromagnetic field, i.e.,
\begin{eqnarray}
&&R_\mu(\omega\in R^+,p_y,p_z,s)=
\frac{1}{\sqrt{2\sinh(\pi\omega})}\nonumber\\
&&[e^{(\frac{\pi\omega}{2})}A_\mu(\omega,p_y,p_z,s)-e^{(-\frac{\pi\omega}{2})}\bar{A}_\mu(-\omega,-p_y,-p_z,s)],\nonumber\\
\\
&&L_\mu(\omega\in R^+,p_y,p_z,s)=\frac{1}{\sqrt{2\sinh(\pi\omega})}\nonumber\\
&&[e^{(\frac{\pi\omega}{2})}A_\mu(-\omega,p_y,p_z,s)-e^{(-\frac{\pi\omega}{2})}\bar{A}_\mu(\omega,-p_y,-p_z,s)],\nonumber\\
\end{eqnarray}
where $R_\mu$ vanishes in the $L$ sector, and $L_\mu$ vanishes in
the $R$ sector. It is noteworthy that $R_\mu(\omega\in
R^+,p_y,p_z,s)$($L_\mu(\omega\in R^+,p_y,p_z,s)$) is the
simultaneous eigenstate of energy, transverse momentum, and helicity
operators with eigenvalues of $\{a\omega,p_y,p_z,s\}$ detected by an
observer with uniform acceleration $a$ in the $R$($L$) Rindler
spacetime\cite{HQZ,Ashtekar}. Moreover, with respect to the inner
product (\ref{inner}), Unruh basis is orthonormal, i.e.,
\begin{eqnarray}
&&(R(\omega,p_y,p_z,s),R(\omega',p'_y,p'_z,s'))=\nonumber\\
&&\delta(\omega-\omega')\delta(p_y-p'_y)\delta(p_z-p'_z)\delta_{ss'},\\
&&(L(\omega,p_y,p_z,s),L(\omega',p'_y,p'_z,s'))=\nonumber\\
&&\delta(\omega-\omega')\delta(p_y-p'_y)\delta(p_z-p'_z)\delta_{ss'},\\
&&(R(\omega,p_y,p_z,s),L(\omega',p'_y,p'_z,s'))=0.
\end{eqnarray}
Whence the quantum electromagnetic field can be reformulated as
\begin{eqnarray}
\hat{A}_\mu=&&\int_0^\infty d\omega\int_{-\infty}^\infty
dp_y\int_{-\infty}^\infty dp_z\sum_{s=\pm1}\nonumber\\
&&[r(\omega,p_y,p_z,s)R_\mu(\omega,p_y,p_z,s)\nonumber\\
&&+r^\dagger(\omega,p_y,p_z,s)\bar{R}_\mu(\omega,p_y,p_z,s)\nonumber\\
&&+l(\omega,p_y,p_z,s)L_\mu(\omega,p_y,p_z,s)\nonumber\\
&&+l^\dagger(\omega,p_y,p_z,s)\bar{L}_\mu(\omega,p_y,p_z,s)].
\end{eqnarray}
Here $r$ and $r^\dagger$ are the corresponding annihilation and
creation operators for the $R$ Rindler spacetime; similarly, $l$ and
$l^\dagger$ are the corresponding annihilation and creation
operators for the $L$ Rindler spacetime. They satisfy the ordinary
commutation relations as $c$ and $c^\dagger$ do. Furthermore, they
can be related to $c$ and $c^\dagger$ by Bogoliubov transformation,
i.e.,
\begin{eqnarray}
&&r(\omega,p_y,p_z,s)=
\frac{1}{\sqrt{2\sinh(\pi\omega})}\nonumber\\
&&[e^{(\frac{\pi\omega}{2})}c(\omega,p_y,p_z,s)+e^{(-\frac{\pi\omega}{2})}c^\dagger(-\omega,-p_y,-p_z,s)],\nonumber\\
\\
&&l(\omega,p_y,p_z,s)=\frac{1}{\sqrt{2\sinh(\pi\omega})}\nonumber\\
&&[e^{(\frac{\pi\omega}{2})}c(-\omega,p_y,p_z,s)+e^{(-\frac{\pi\omega}{2})}c^\dagger(\omega,-p_y,-p_z,s)];\nonumber\\
\end{eqnarray}
or vice versa
\begin{eqnarray}
&&c(\omega,p_y,p_z,s)=
\frac{1}{\sqrt{2\sinh(\pi\omega})}\nonumber\\
&&[e^{(\frac{\pi\omega}{2})}r(\omega,p_y,p_z,s)-e^{(-\frac{\pi\omega}{2})}l^\dagger(\omega,-p_y,-p_z,s)],\nonumber\\
\\
&&c(-\omega,p_y,p_z,s)=\frac{1}{\sqrt{2\sinh(\pi\omega})}\nonumber\\
&&[e^{(\frac{\pi\omega}{2})}l(\omega,p_y,p_z,s)-e^{(-\frac{\pi\omega}{2})}r^\dagger(\omega,-p_y,-p_z,s)].\nonumber\\
\end{eqnarray}
Note that the vacuum state killed by the annihilation operator $c$
is equivalent to the ordinary Minkowski one\cite{Colosi}. Hence one
obtains the expression for the ordinary Minkowski vacuum in the mode
$A_\mu(\omega,p_y,p_z,s)$ as a Rindler state, i.e.,
\begin{eqnarray}
|0\rangle^M_{\omega,p_y,p_z,s}=&&\sqrt{\frac{2\sinh(\pi\omega)}{e^{(\pi\omega)}}}\sum_{n=0}^\infty
e^{(-n\pi \omega)}
\nonumber\\
&&|n(\omega,p_y,p_z,s)\rangle^R\otimes|n(\omega,-p_y,-p_z,s)\rangle^L,\nonumber\\
\end{eqnarray}
where
$|n(\omega,p_y,p_z,s)\rangle^R$($|n(\omega,p_y,p_z,s)\rangle^L$)
denotes the state with $n$ particles in Unruh mode
$R_\mu(\omega,p_y,p_z,s)$($L_\mu(\omega,p_y,p_z,s)$). Furthermore,
we have
\begin{eqnarray}
&&|1\rangle^M_{\omega,p_y,p_z,s}=c^\dagger(\omega,p_y,p_z,s)|0\rangle^M=\nonumber\\
&&[1-e^{(-2\pi\omega)}]\sum_{n=0}^\infty e^{-n\pi\omega}\sqrt{n+1}\nonumber\\
&&|(n+1)(\omega,p_y,p_z,s)\rangle^R\otimes|n(\omega,-p_y,-p_z,s)\rangle^L\nonumber\\
&&\prod_{\{\omega',p'_y,p'_z,s'\}\neq\{\omega,p_y,p_z,s\}}|0\rangle^M_{\omega',p'_y,p'_z,s'}.\nonumber\\
\label{expansion}
\end{eqnarray}

\section{Entanglement for Electromagnetic Fields in Non-inertial Reference Frames\label{QE}}
In order to analyze quantum entanglement for electromagnetic field
in non-inertial reference frames, firstly following previous work
\cite{AM1,AM2,FM,AFMT}, we can also take into account the particle
number entangled state in the inertial reference frame associated
with Alice, i.e.,
\begin{equation}
|\varphi\rangle=\frac{1}{\sqrt{2}}(|0\rangle^M_A|0\rangle^M_B+|1\rangle^M_A|1\rangle^M_B).
\end{equation}
It is easy to show that the helicity structure of photon has no
influence in this case, and the corresponding calculation goes
straightforward, exactly the same as that for scalar particle, which
thus justifies modeling photon with scalar particle in investigation
of quantum entanglement in non-inertial reference frames for the
particle number entangled state\cite{AM1,AM2,FM}.

We would next like to concentrate onto two photons' maximally
helicity entangled state in the inertial reference frame, i.e.,
\begin{eqnarray}
|\psi\rangle&=&\frac{1}{\sqrt{2}}(|1\rangle^M_{\omega,p_y,p_z,1A}|1\rangle^M_{\omega,-p_y,-p_z,-1B}\nonumber\\
&&+|1\rangle^M_{\omega,p_y,p_z,-1A}|1\rangle^M_{\omega,-p_y,-p_z,1B}),\label{helicity}
\end{eqnarray}
which also seems to be more popular than the particle number
entangled state in quantum information science. For later
convenience, we shall rewrite (\ref{helicity}) as
\begin{equation}
|\psi\rangle=\frac{1}{\sqrt{2}}(|1\rangle^M_{+\uparrow
A}|1\rangle^M_{-\downarrow B} +|1\rangle^M_{+\downarrow
A}|1\rangle^M_{-\uparrow B}).
\end{equation}
To describe this state from the viewpoint of the non-inertial
observer Bob, firstly we shall employ (\ref{expansion}) to expand
this state. Since Bob is causally disconnected from the $L$ sector,
we must take trace over all of the $L$ sector modes, which results
in a mixed density matrix between Alice and Bob, i.e.,
\begin{eqnarray}
\rho_{AB}=&&\frac{[1-e^{(-\frac{2\pi E}{a})}]^2}{2}\sum_{n=0}^\infty e^{(-\frac{2n\pi E}{a})}(n+1)\nonumber\\
&&(\ \ |1\rangle^M_{+\uparrow A}|n+1\rangle^R_{-\downarrow B}\langle1|^M_{+\uparrow A}\langle n+1|^R_{-\downarrow B}\nonumber\\
&&+|1\rangle^M_{+\uparrow A}|n+1\rangle^R_{-\downarrow B}\langle1|^M_{+\downarrow A}\langle n+1|^R_{-\uparrow B}\nonumber\\
&&+|1\rangle^M_{+\downarrow A}|n+1\rangle^R_{-\uparrow B}\langle1|^M_{+\uparrow A}\langle n+1|^R_{-\downarrow B}\nonumber\\
&&+|1\rangle^M_{+\downarrow A}|n+1\rangle^R_{-\uparrow
B}\langle1|^M_{+\downarrow A}\langle n+1|^R_{-\uparrow B}\
),\nonumber\\\label{joint}
\end{eqnarray}
where $a$ denotes Bob's acceleration, and $E=a\omega$ is the energy
sensitive to Bob's detector.

To determine whether this mixed state is entangled or not, we here
use the partial transpose criterion\cite{Peres2}. It states that if
the partial transposed density matrix of a system has at least one
negative eigenvalue, it must be entangled, otherwise it has no
distillable entanglement, but may have other types of entanglement.
After a straightforward calculation, the partial transposed density
matrix can be obtained as
\begin{eqnarray}
\rho_{AB}^T=&&\frac{[1-e^{(-\frac{2\pi E}{a})}]^2}{2}\sum_{n=0}^\infty e^{(-\frac{2n\pi E}{a})}(n+1)\nonumber\\
&&(\ \ |1\rangle^M_{+\uparrow A}|n+1\rangle^R_{-\downarrow B}\langle1|^M_{+\uparrow A}\langle n+1|^R_{-\downarrow B}\nonumber\\
&&+|1\rangle^M_{+\downarrow A}|n+1\rangle^R_{-\downarrow B}\langle1|^M_{+\uparrow A}\langle n+1|^R_{-\uparrow B}\nonumber\\
&&+|1\rangle^M_{+\uparrow A}|n+1\rangle^R_{-\uparrow B}\langle1|^M_{+\downarrow A}\langle n+1|^R_{-\downarrow B}\nonumber\\
&&+|1\rangle^M_{+\downarrow A}|n+1\rangle^R_{-\uparrow B}\langle1|^M_{+\downarrow A}\langle n+1|^R_{-\uparrow B}\ ),\nonumber\\
\end{eqnarray}
whose eigenvalues are easy to be computed, specifically those
belonging to the \textit{n}th diagonal block are
$\frac{[1-e^{[-\frac{2\pi E}{a})}]^2}{2}e^{(-\frac{2n\pi
E}{a})}(n+1)(1,1,1,-1)$. Thus the state as seen by Bob will be
always entangled if only the acceleration is finite. However,
quantification of the distillable entanglement can not be carried
out in this case. Therefore we only provide an upper bound of the
distillable entanglement by the logarithmic negativity\cite{VW}. It
is defined as $N(\rho)=\log_2||\rho^T||_1$, where$||\ ||_1 $is the
trace norm of a matrix. Whence the logarithmic negativity is given
by
\begin{equation}
N(\rho_{AB})=\log_2\{2[1-e^{(-\frac{2\pi E}{a})}]^2\sum_{n=0}^\infty
e^{(-\frac{2n\pi E}{a})}(n+1)\}=1,
\end{equation}
which is independent of the acceleration of Bob.

Further, we can also make an estimation of the total correlation in
the state by employing the mutual information, i.e.,
$I(\rho_{AB})=S(\rho_A)+S(\rho_B)-S(\rho_{AB})$ where
$S(\rho)=-Tr(\rho\log_2\rho)$ is the entropy of the matrix $\rho$.
According to (\ref{joint}), the entropy of the joint state reads
\begin{eqnarray}
S(\rho_{AB})&&=-[1-e^{(-\frac{2\pi E}{a})}]^2\sum_{n=0}^\infty e^{(-\frac{2n\pi E}{a})}(n+1)\nonumber\\
&&\log_2\{[1-e^{(-\frac{2\pi E}{a})}]^2e^{(-\frac{2n\pi
E}{a})}(n+1)\}.
\end{eqnarray}
Tracing over Alice's states yields Bob's density matrix as
\begin{eqnarray}
\rho_B=&&\frac{[1-e^{(-\frac{2\pi E}{a})}]^2}{2}\sum_{n=0}^\infty
e^{(-\frac{2n\pi E}{a})}(n+1)\nonumber\\
&&(\ |n+1\rangle^R_{-\downarrow B}\langle n+1|^R_{-\downarrow
B}+|n+1\rangle^R_{-\uparrow B}\langle
n+1|^R_{-\uparrow B}),\nonumber\\
\end{eqnarray}
whose entropy is
\begin{eqnarray}
S(\rho_B)=&&1-[1-e^{(-\frac{2\pi E}{a})}]^2\sum_{n=0}^\infty e^{(-\frac{2n\pi E}{a})}(n+1)\nonumber\\
&&\log_2\{[1-e^{(-\frac{2\pi E}{a})}]^2e^{(-\frac{2n\pi
E}{a})}(n+1)\}.
\end{eqnarray}
Similarly, tracing over Bob's states, we obtain Alice's density
matrix as
\begin{equation}
\rho_A=\frac{1}{2}(|1\rangle^M_{+\uparrow A}\langle1|^M_{+\uparrow
A}+|1\rangle^M_{+\downarrow A}\langle1|^M_{+\downarrow A}),
\end{equation}
which has an entropy $S(\rho_A)=1$. As a result, the mutual
information is $I(\rho_{AB})=2$, which is the same for any uniformly
accelerated observer, no matter how much the magnitude of
acceleration is.

Therefore, as seen by Bob, the helicity entanglement in non-inertial
reference frames shows a remarkably interesting behavior, which is
obviously different from the case for the particle number
entanglement. In particular, the calculable logarithmic negativity
and mutual information both remain constant for the photon helicity
entangled state, which is in strong contrast to the particle number
entangled state, where they both degrade with the increase of
acceleration. All of this seems to imply that the photon helicity
entangled state is more robust against the perturbation of
acceleration or gravitation than the particle number entangled
state, thus can be used as a more effective resource for performing
some quantum information processing technology.
\section{Conclusions and Discussions}
In this paper we have attempted to provide an analysis of quantum
entanglement of electromagnetic field in non-inertial reference
frames. In particular, we find that the maximally helicity entangled
state is a stable state under acceleration in the sense of its
logarithmic negativity and mutual information, which is obviously a
novel result, completely different from the case for the particle
number entangled state.

As is mentioned in the beginning, the major difference between our
work and previous ones concerning quantum entanglement in
non-inertial frames is that we have considered the helicity
entanglement while previous ones only focus on the entanglement in
particle number. The helicity structure is special to photons, which
is a completely new trait that can not be presented in the case of
scalar particles. It is tempting to say that the entanglement of the
discrete degrees of freedom is generally different from the particle
number entanglement. Especially, the entangled state seems more
immune to the destruction of the acceleration or gravitation in
discrete degrees of freedom than particle number. To confirm this
conjecture, the spin entanglement of Dirac field in non-inertial
reference frames is a necessary and important task worthy of further
investigation. Since Dirac particle is constrained by Pauli
exclusion principle, it is a qubit-qubit system and the evaluation
of the corresponding entanglement is much easier, especially the
entanglement of formation can be explicitly
calculated\cite{Wootters}. Such a detailed analysis of the spin
entanglement in non-inertial reference frames and related problems
is expected to be reported elsewhere.
\section*{Acknowledgements}
We would like to give much gratitude to Chopin Soo for his
stimulating suggestion on this work and Bo Hu for his figure plotted
here. In addition, we gratefully acknowledge Steven J. van Enk for
his insightful and helpful comments. Valuable discussions from Paul
Alsing, Robert Mann, and Tracy Tessier are also much appreciated. Y.
Ling's work is partly supported by NSFC(Nos.10205002 and 10405027)
and SRF for ROCS. S. He's work is supported by NSFC(Nos.10235040 and
10421003). W. Qiu's work is supported by NSFC(No.10547116), the
Science Research Fund of Huzhou Teachers College(No.KX21001) and the
Science Research Fund of Huzhou City(No.KY21022). H. Zhang's work is
supported in part by NSFC(Nos.10373003 and 10533010).


\begin{thebibliography}{99}
\bibitem{BEZ}The Physics of Quantum Information,
Springer-Verlag, 2000, D. Bouwmeester \emph{et al.}(Eds.).
\bibitem{PV}M. B. Plenio and S. Virmani, quant-ph/0504163.
\bibitem{AET}S. J. van Enk and T. Rudolph, Quant. Inf. Comput. 3: 423(2003).
\bibitem{PT}A. Peres and D. R. Terno, Rev. Mod. Phys. 76: 93(2004).
\bibitem{Shi}Y. Shi, Phys. Rev. D70: 105001(2004).
\bibitem{Peres1}A. Peres \emph{et al.}, Phys. Rev. Lett. 88:
230402(2002).
\bibitem{AM1}P. M. Alsing and G. J. Milburn, Quant. Inf. Comput. 2: 487(2002).
\bibitem{GA}R. M. Gingrich and C. Adami, Phys. Rev. Lett. 89:
270402(2002).
\bibitem{He}S. He \emph{et al.}, quant-ph/0701233.
\bibitem{AM2}P. M. Alsing and G. J. Milburn, Phys. Rev. Lett. 91:
180404(2003).
\bibitem{AM3}P. M. Alsing \emph{et al.}, J. Optics. B6: S834
(2004).
\bibitem{FM}I. Fuentes-Schuller and R. B. Mann, Phys. Rev. Lett. 95:
120404(2005).
\bibitem{AFMT}P. M. Alsing \emph{et al.}, Phys. Rev. A74: 032326(2006).
\bibitem{Moretti}V. Moretti, J. Math. Phys. 38: 2922(1997).
\bibitem{HQZ}Y. Hu \emph{et al.}, J. Math. Phys. 47: 052304(2006).
\bibitem{Colosi}D. Colosi, Nuovo Cim. B115: 1101(2000).
\bibitem{Ashtekar}A. Ashtekar, J. Math. Phys. 27: 824(1986).
\bibitem{Peres2}A. Peres, Phys. Rev. Lett. 77: 1413(1996).
\bibitem{VW}G. Vidal and R. F. Werner, Phys. Rev. A65: 032314(2002).
\bibitem{Wootters}W. K. Wootters, Phys. Rev. Lett. 80: 2245(1998).
\end{thebibliography}
\end{document}